\documentstyle[12pt]{article} 
\newcommand{\beq}{\begin{equation}}
\newcommand{\eeq}{\end{equation}}
\newcommand{\beqa}{\begin{eqnarray}}
\newcommand{\eeqa}{\end{eqnarray}}
\newcommand{\ba}{\begin{array}}
\newcommand{\ea}{\end{array}}

\begin{document}

\begin{center}
{\Large \bf BEC in Nonextensive Statistical Mechanics} 
\vskip 0.8cm 
Luca Salasnich \\
\vskip 0.5cm
Istituto Nazionale per la Fisica della Materia, Unit\`a di Milano,\\ 
Dipartimento di Fisica, Universit\`a di Milano, \\ 
Via Celoria 16, 20133 Milano, Italy 
\end{center}

\vskip 1.5cm

\begin{center}
{\bf Abstract} 
\end{center}
We discuss the Bose-Einstein condensation (BEC) for an ideal gas 
of bosons in the framework of Tsallis's nonextensive 
statistical mechanics. 
We study the corrections to the standard BEC formulas 
due to a weak nonextensivity of the system. 
In particular, we consider three cases 
in the D-dimensional space: 
the homogeneous gas, the gas in a harmonic trap and 
the relativistic homogenous gas. 
The results show that small deviations from 
the extensive Bose statistics produce remarkably large 
changes in the BEC transition temperature.  

\vskip 0.5cm


\newpage

A decade ago, Tsallis introduced a nonextensive statistical mechanics 
(NSM) to describe systems for which the additivity property of entropy 
does not hold.$^{1}$ The NSM can describe systems for which 
long-range microscopic memory, fractal space-time constraints 
or long-range interactions affect the thermalization process.$^{2}$ 
The NSM is characterized by a parameter $q$ such that 
$(q-1)$ is a measure of the lack of extensivity: in the limit 
$q\to 1$ one recovers the familiar statistical mechanics but for 
$q\ne 1$ one obtains generalized 
Boltzmann, Fermi and Bose distributions.$^{3}$
In the last few years the NSM has been applied in different contexts 
like solar neutrinos,$^{4}$ high energy nuclear collisions$^{5}$ and 
the cosmic microwave background radiation.$^{6}$ 
In such cases it has been found that a small deviation 
from standard statistics is sufficient 
for eliminating the discrepancy between 
theoretical calculations and experimental data.  
\par 
Recently, there has been a renewed theoretical interest on Bose-Einstein 
condensation (BEC) (for a review see Ref. 7), 
motivated by the experimental achievement of BEC 
with trapped weakly-interacting alkali-metal atoms.$^{8}$ 
In this paper we analyze the consequences of weak nonextensivity 
on BEC for an ideal Bose gas. From the generalized Bose-Einstein 
distribution we derive the BEC transition temperature, 
the condensed fraction and the energy per particle 
in three different cases: the homogeneous gas, the gas in a harmonic trap and 
the relativistic homogenous gas. All the calculations are performed 
by assuming a D-dimensional space. 
\par
For a quantum gas of identical bosons in the grand canonical ensemble, 
the NSM predicts that the average number of particles 
with energy $\epsilon$ is given by 
\beq 
\langle n(\epsilon )\rangle_q = {1\over \left[1+ 
\beta (q-1) (\epsilon - \mu)\right]^{1/(q-1)} - 1} \; ,
\eeq
where $\mu$ is the chemical potential and $\beta=1/(kT)$ 
with $k$ the Boltzmann constant and $T$ the temperature.$^{2}$
This generalized distribution follows from the minimization of the 
Tsallis's generalized entropy under 
the dilute gas assumption, namely the different single-particle 
states of the systems are regarded as independent. 
Thus, this is not an exact formula 
but it has been shown to be extremely accurate, 
in particular near $q=1$.$^{9}$
When $q<1$ the generalized distribution has an upper cut-off: 
$(\epsilon - \mu ) \le kT/(1-q)$. In the limit $q\to 1$ 
the generalized distribution becomes the standard Bose-Einstein 
distribution. For $q>1$ there is no cut-off and the (power-law) 
decay is slower than exponential. 
Because of the unphysical cut-off for $q<1$, in this paper 
we discuss only the case $q\ge 1$. 
\par 
We want study the effects of weak nonextensivity on the BEC 
properties. We assume that $(q-1)<1$ and 
by performing a Taylor expansion of the generalized Bose distributions 
in the parameter $(q-1)$, at first order we obtain 
\beq
\langle n(\epsilon )\rangle_q = 
{1\over e^{\beta(\epsilon -\mu)}-1}  
+ {1\over 2} (q-1) {\beta^2 (\epsilon - \mu)^2 e^{\beta(\epsilon -\mu)} 
\over (e^{\beta(\epsilon -\mu)}-1)^2 } \; .
\eeq
This is the weak nonextensivity correction to the standard 
Bose-Einstein distribution and the starting point for our calculations. 
\par
The total number of particle for our system of non-interacting 
bosons reads
\beq
N=\int_0^{\infty} 
d\epsilon \; \rho(\epsilon ) \; \langle n(\epsilon )\rangle_q \; ,
\eeq 
where $\rho (\epsilon )$ is the density of states. 
It can be obtained from the formula 
\beq
\rho(\epsilon ) = \int {d^D{\bf q} d^D{\bf p}\over (2\pi\hbar)^D} 
\delta (\epsilon - H({\bf p},{\bf q})) \; ,  
\eeq
where $H({\bf p},{\bf q})$ is the classical single-particle 
Hamiltonian of the system in a D-dimensional space. 
It is easy to show that for a homogenous gas 
the density of states in a D-dimensional box of volume $V$ is given by 
\beq 
\rho(\epsilon ) = {V\over \Gamma(D/2)} 
\left({m\over 2\pi \hbar^2}\right)^{D/2} \epsilon^{(D-2)/2} \; ,  
\eeq
where $m$ is the mass of the particle. 
Instead, for a gas in a harmonic trap one finds  
\beq
\rho(\epsilon ) = {\epsilon^{D-1}\over 
(\hbar \bar{\omega})^D \Gamma(D)} \; ,
\eeq 
where $\bar{\omega}$ is the geometric average of the trap frequencies. 
$\Gamma(x)$ is the factorial function. 
\par
At the BEC transition temperature $T_q$, 
the chemical potential $\mu$ is zero and 
at $\mu=0$ the number of particles $N$ 
can be analytically determined from Eq. (2) and (3). 
By inverting the function $N=N(T_q)$ one finds 
the transition temperature. It is given by 
\beq
kT_q = \left({2\pi \hbar^2 \over m}\right) 
{(N/V)^{2/D}\over \zeta (D/2)^{2/D}} 
\left[1 + {1\over 2}(q-1) 
{\Gamma(D/2+2) \zeta(D/2+1)\over \Gamma(D/2) \zeta(D/2) }
\right]^{-2/D} \; 
\eeq 
for the homogenous gas, and by 
\beq
kT_q ={\hbar \bar{\omega}\over \zeta (D)^{1/D}} N^{1/D} 
\left[1 + {1\over 2}(q-1){\Gamma(D+2)\zeta(D+1)\over \Gamma(D)\zeta(D)} 
\right]^{-1/D} \;  
\eeq
for a gas in a harmonic trap. 
$\zeta(x)$ is the Riemann $\zeta$-function. 
Obviously, for $q=1$ one recovers 
standard BEC formulas. Moreover one observes that for $D=2$ 
there is no BEC in the homogenous gas because $\zeta(1)=\infty$.  
Instead, BEC is possible with $D=2$ in a harmonic trap. 
Note that the inclusion of an attractive interaction can modify 
the stability of the Bose condensate. 
A discussion of the the role of dimensionality 
in the stability of a weakly-interacting condensate 
can be found in Ref. 10. 
\par
An inspection of Eq. (7) and (8) shows that the critical temperature 
$T_q$ grows by increasing the nonextensive parameter $q$ and 
the space dimension $D$. It is important to stress that such effect is 
quite strong. For example, with $q=1.1$ and $D=3$ we have that 
the relative difference $(T_q-T_1)/T_1$ is $6.32\%$ 
for the homogenous gas and and $15.48\%$ for the gas 
in a harmonic trap. 
\par
Below $T_q$, a macroscopic number $N_0$ of particle occupies 
the single-particle ground-state of the system. It follows 
that Eq. (3) gives the number $N-N_0$ of non-condensed particles and 
the condensed fraction is $N_0/N=1-(T/T_q)^{D/2}$ for the homogenous 
gas and $N_0/N=1-(T/T_q)^D$ for the gas in harmonic trap. 
For the sake of completeness, we calculate also the energy 
\beq
E= \int_0^{\infty} d\epsilon \; 
\epsilon \; \rho(\epsilon ) \; \langle n(\epsilon )\rangle_q \; . 
\eeq 
From the energy one can easily obtain the specific heat and the other 
thermodynamical quantities. We find 
\beq 
{E\over KT} = V \left({kT\over 2\pi\hbar^2}\right)^{D/2} {D\over 2} 
\zeta(D/2+1) 
\left[1 + {1\over 2}(q-1) 
{\Gamma(D/2+3) \zeta(D/2+2)\over \Gamma(D/2+1) \zeta(D/2+1) }
\right] \; 
\eeq 
for the homogenous gas, and by 
\beq
{E\over KT} = \left({kT\over \hbar \bar{\omega}}\right)^D D \zeta(D+1) 
\left[1 + {1\over 2}(q-1) 
{\Gamma(D+3) \zeta(D+2)\over \Gamma(D+1) \zeta(D+1) }
\right] \; 
\eeq
for a gas in a harmonic trap. Note that our formulas of the energy can be 
easily generalized above the critical temperature $T_q$ by substituting 
the Riemann function $\zeta(D)$ with the polylogarithm 
function $Li_{D}(z)=\sum_{k=1}^{\infty}z^k/k^D$, that depends 
on the fugacity $z=e^{\beta \mu}$. 
\par
In the case of a relativistic gas, the total number of particles is not 
conserved because of the production 
of antiparticles, which becomes relevant when $kT$ is comparable 
with $mc^2$. The conserved quantity is the difference 
between the number $N$ of particles and the number $\bar{N}$ of 
antiparticles, i.e. the net conserved {\it charge} 
\beq
Q=N-\bar{N}=
\int d\epsilon \; \rho(\epsilon ) 
\left[ \langle n(\epsilon )\rangle_q - 
\langle \bar{n}(\epsilon )\rangle_q \right] \; ,
\eeq 
where $\langle \bar{n}(\epsilon )\rangle_q$ is obtained 
from $\langle n(\epsilon )\rangle_q$ with the substitution 
$\mu \to - \mu$. Thus the chemical potential $\mu$ 
describes both bosons and antibosons: the sign of $\mu$ 
indicates whether particles outnumber antiparticles or vice. 
Moreover, because both $\langle n(\epsilon )\rangle_q$ and 
$\langle \bar{n}(\epsilon )\rangle_q$ must be positive definite, 
it follows that $|\mu |\le mc^2$.$^{11}$ 
\par
As well known, the classical 
single-particle Hamiltonian of a relativistic ideal gas is 
$H=\sqrt{p^2c^2 + m^2c^4}$ and the density of states reads 
\beq
\rho(\epsilon )={V 2\pi^{D/2} \over (2\pi \hbar c)^D \Gamma(D/2)} 
\epsilon (\epsilon^2-m^2 c^4)^{(D-2)/2} \; . 
\eeq
It is interesting to observe that 
in the ultrarelativistic limit, the density of states is 
$\rho(\epsilon )=(V 2\pi^{D/2})/((2\pi \hbar c)^D \Gamma(D/2)) 
\epsilon^{(D-1)}$ and it has the same power law of the 
non-relativistic gas in a harmonic potential. 
The critical temperature $T_q$ at which BEC occurs corresponds 
to $|\mu| = mc^2$. In the ultrarelativistic region $kT \gg mc^2$ 
one can expand $Q$ at first order in $\mu$ and then obtains 
$$
kT_q = \left( { (2\pi \hbar c)^D \Gamma(D/2)  
\over 4\pi^{D/2} \Gamma(D) \zeta(D-1)} 
{|Q|/V \over mc^2 } \right)^{1/(D-1)} \times 
$$
\beq
\times \left[1 + {1\over 2}(q-1) 
{(D-1)\Gamma(D+1)\zeta(D) 
\over \Gamma(D)\zeta(D-1)} \right]^{-1/(D-1)} \; .
\eeq
Note that, as in the non-relativistic case, for a homogenous gas 
there is BEC only for $D>2$. Also for the relativistic gas the 
critical temperature $T_q$ is a growing function of the nonextensive 
parameter $q$ (for $q\ge 1$) and of the space dimension $D$. 
By using the previously introduced 
values $q=1.1$ and $D=3$ we find $(T_q-T_1)/T_1=6.83\%$. 
Finally, we obtain that below $T_q$ 
the condensed fraction reads $Q_0/Q=1-(T/T_q)^{(D-1)}$.  
\par
In conclusion, we have analyzed the consequences of 
Tsallis's nonextensive statistical mechanics on BEC. 
We have studied three non-interacting systems with a generic 
spatial dimension: the homogeneous gas, the gas in a harmonic trap and 
the relativistic homogenous gas. 
The calculations show that a very small deviation from 
the extensive Bose statistics produces remarkable changes 
in the BEC transition temperature. 
This result may have important consequences, for instance 
in the formation of Quark-Gluon Plasma$^{12}$ 
and in the thermodynamics of the Higgs field in the early Universe.$^{13}$ 
We observe that the inter-particle interaction can strongly modify 
the BEC transition temperature and the condensate properties:  
one of our future projects will be the study of nonextensive 
statistical mechanics for interacting systems. 

\newpage 

\section*{References}

\begin{description}

\item{\ 1.} C. Tsallis, J. Stat. Phys. {\bf 52}, 479 (1988). 

\item{\ 2.} E.M.F. Curado and C. Tsallis, J. Phys. A {\bf 24}, L69 (1991). 

\item{\ 3.} F. Buyukkilic, D. Demirhan, and A. Gulec, Phys. Lett. A 
{\bf 197}, 209 (1995). 

\item{\ 4.} G. Kaniadakis, A. Lavagno, and P. Quarati, Phys. Lett. B 
{\bf 369}, 308 (1996). 

\item{\ 5.} G. Kaniadakis, A. Lavagno, M. Lissia, and P. Quarati, 
in Proceedings of the VII Workshop on 
{\it Perspectives on Theoretical Nuclear Physics}, pp. 293, 
Ed. A. Fabrocini, G. Pisent and S. Rosati (Edizioni ETS, Pisa, 1999). 

\item{\ 6.} C. Tsallis, F.C.S. Barreto, and E.D. Loh, Phys. Rev. 
E {\bf 52}, 1447 (1995).

\item{\ 7.} F. Dalfovo, S. Giorgini, L.P. Pitaevskii, and S. 
Stringari, Rev. Mod. Phys. {\bf 71}, 463 (1999). 

\item{\ 8.} M.H. Anderson, {\it et al.}, Science {\bf 269}, 189 (1995); 
K.B. Davis, {\it et al.}, Phys. Rev. Lett. {\bf 75}, 
3969 (1995); C.C. Bradley, {\it et al.}, 
Phys. Rev. Lett. {\bf 75}, 1687 (1995). 

\item{\ 9.} Q.A. Wang and A. Le Mehaute, Phys. Lett. A {\bf 235}, 
222 (1997). 

\item{\ 10.} L. Salasnich, Mod. Phys. Lett. B {\bf 11}, 1249 (1997); 
{\bf 12}, 649 (1998). 

\item{\ 11.} H.E. Haber and H.A. Weldon, Phys. Rev. Lett. {\bf 23}, 
1497 (1981); J.I. Kapusta, Phys. Rev. D {\bf 24}, 426 (1981). 

\item{\ 12.} B. Muller, {\it The Physics of Quark-Gluon Plasma} 
(Springer, Berlin, 1985). 

\item{\ 13.} A.D. Linde, {\it Particle Physics and Inflationary 
Cosmology} (Harwood Academic, London, 1988). 

\end{description}

\end{document}